\documentstyle[12pt]{article}

\begin{document}
\input{psfig.sty}
\begin{flushright}
\baselineskip=12pt
MADPH-99-1150 \\
\end{flushright}

\begin{center}
\vglue 1.5cm
{\Large\bf Classification of 5-Dimensional Space-Time with Parallel 3-Branes\\}
\vglue 2.0cm
{\Large Tianjun Li~\footnote{E-mail: li@pheno.physics.wisc.edu,
phone: (608) 262-9820, fax: (608) 262-8628.}}
\vglue 1cm
\begin{flushleft}
Department of Physics, University of Wisconsin, Madison, WI 53706,  
U.  S.  A.
\end{flushleft}
\end{center}

\vglue 1.5cm
\begin{abstract}
If the fifth dimension is one-dimensional connected manifold, up to diffeomorphic, the only 
possible
space-time will be $M^4 \times R^1$, $M^4 \times R^1/Z_2$,
$M^4 \times S^1$ and $M^4 \times S^1/Z_2$. 
And there exist two 
possibilities on cosmology constant along the fifth dimension: the cosmology
constant is constant, and the cosmology constant
is sectional constant.
We construct the general models with  parallel 3-branes
and with constant/sectional constant cosmology constant along the
fifth dimension on
those kinds of the space-time, and  point out that for compact fifth dimension, the sum of 
the brane tensions
is zero, for non-compact fifth dimension, the sum of the brane tensions is positive.
We assume the observable brane which includes our world should have
positive tension, and obtain that the gauge hierarchy problem can be solved in those
scenarios. We also discuss some simple models.
\\[1ex]
\\[1ex]

\end{abstract}

\vspace{0.5cm}
\begin{flushleft}
\baselineskip=12pt
December 1999\\
\end{flushleft}
\newpage
\setcounter{page}{1}
\pagestyle{plain}
\baselineskip=14pt

\section{Introduction}

Experiments at LEP and Tevatron have given the strong
support to the Standard Model $SU(3)_c \times SU(2)_L \times U(1)$.
 However, the Standard Model has some unattractive features which
may imply the new physics. 
One of these problems is that the gauge forces and the gravitational force
are not unified. Another is
the gauge hierarchy problem between the weak scale and the 4-dimensional
Planck scale. Previously, two solutions to the
gauge hierarchy problem have
been proposed: one is the idea of the technicolor and  compositeness
which lacks calculability, and the other is the idea
of supersymmetry.    

More than one year ago, it was suggested that the large 
compactified extra dimensions may also be the solution to the
gauge hierarchy problem~\cite{AADD}. Furthermore,
about half year ago, Randall and Sundrum~\cite{LRRS} proposed another scenario
that the extra dimension is an orbifold,
and  the size of the extra dimension is not large
but the 4-dimensional mass scale in the standard model is
suppressed by an exponential factor from 5-dimensional mass
scale. In addition,  they suggested that
the fifth dimension might be infinity~\cite{LRRSN}, and there may exist only one
brane with positive tension at the origin, but, there  exists the  gauge
hierarchy problem.  The remarkable aspect of
the second scenario is that it gives rise to a localized graviton field.
Combining those results, Lykken and Randall obtained the following
physical picture~\cite{JLLR}: the graviton is localized on Planck brane, we
live on a brane separated from the Planck brane about 30 Planck lengths 
along the fifth dimension. 
 On our brane, the mass scale
in the Standard Model is suppressed
exponentially, which gives the low energy scale.
We generalized those scenarios and obtained the scenario with the following
property~\cite{LTJ}: the 4-dimensional Planck scale is generated from the low 5-dimensional
Planck scale by an exponential hierarchy, and the mass scale in the
Standard model, which is contained in the observable brane, is not rescaled.
In short, recently, this kind of compactification or similar idea has attracted
a lot of attentions~\cite{MG, HDDK, MBW, TN, CGKT, NK, HV, IO, ABKS, AK, HBDK, TSMS, 
CVETIC,
LTJ, LTJI, LTJII, KRDDG, DWFGK, CY, CP, GCS, CGRT, CCYS, AEN, ODA, HSTT, NEWNK, SNAM}.
By the way, supergravity domain walls was discussed previously in the 4-dimensional
space-time~\cite{MCHS}. 

In the later model building, there exist several approachs:
(I) The 3-branes are parallel along the fifth dimension
and the observable brane is one of those branes: Oda constructed the
model on the space-time $M^4\times S^1$ where $M^4$ is the four-dimensional
Minkowski space, and  cosmology constant is constant~\cite{IO}. However, his
solution is not general. H. Hatanaka, M. Sakamoto, M. Tachibana, 
and K. Takenaga constructed the model on the space-time $M^4\times S^1$ and 
cosmology constant is sectional constant along the fifth dimension~\cite{HSTT}. And we 
constructed the model
with only parallel positive tension 3-branes on $M^4\times R^1$ and
$M^4\times R^1/Z_2$~\cite{LTJI}. 
(II) The observable sector or our world lives in the brane 
intersection/junction ~\cite{HDDK, CCYS, AEN}.
(III) Many brane intersections/junctions, which might be thought to be
 the combination
of the approach (I) and (II)~\cite{LTJII, NEWNK}.
 In this paper, we classify the models of the
first approach, or in other words, we classify the models on the 5-dimensional space-time 
with 
parallel 3-branes.  From differential topology/manifold, up to
diffeomorphic, there is only one connected non-compact one-dimensional manifold: $R^1$,
there is only one connected non-compact one-dimensional manifold with boundary:
$H_1$ or $R^1/Z_2$ ( the equivalence class is $y \sim -y$ ), there is
only one connected compact one-dimensional manifold $S^1$, and there is
only one connected compact one-dimensional manifold with boundary $S^1/Z_2$ or
$I$.  Therefore, the space-time we consider are $M^4\times R^1$,  
$M^4\times R^1/Z_2$, $M^4\times S^1$,  
$M^4\times S^1/Z_2$.
 Furthermore, there are two possibilities for the cosmology constant along the fifth 
dimension:
  the cosmology
constant is constant, and the cosmology constant
is sectional constant.

Before our discussion, we would like to explain our  
assumption and notation which are similar to those in~\cite{LTJ, LTJI}. 
 We assume that all the gauge forces are unified on each 3-brane
 if there exist gauge forces. The 5-dimensional GUT scale
 on i-th 3-brane  $M5_{GUT}^{(i)}$  and the 5-dimensional Planck scale $M_X$ are defined as
 the GUT scale and Planck scale in the 5-dimensional fundamental
 metric, respectively. 
 The 4-dimensional GUT scale on i-th 3-brane $M_{GUT}^{(i)}$ and the 4-dimenisonal
 Planck scale $M_{pl}$ are defined as the GUT scale and Planck scale in the
 4-dimensional Minkowski Metric ($\eta_{\mu \nu}$). In order to
 avoid the gauge hierarchy problem between the weak scale and
 the 4-dimensional GUT scale $M_{GUT}$ on the observable brane which includes
 our world, we assume the low energy unification~\footnote{We 
 will not explain why $M_{GUT}$ can be at low energy scale here,
but, it is possible if one considers additional particles which 
change the RGE running. Of course, proton decay might be the
problem, but we do not disscuss it here. Moreover, using extra dimensions,
low-scale gauge coupling unification can occur due to power-law
running induced by the presence of Kaluza-Klein states~\cite{KRDDG, GUTP},
and proton-decay constraints might be satisfied in this framework.}. The
 key ansatz is that the 5-dimensional GUT scale on each brane 
 is equal to the 5-dimensional Planck scale.  In addition, as we know, 
an object with negative tension can not be stable, although we might not
need to worry about it if the anti-brane is orientable, so, we consider
the model with positive tension 3-branes and negative tension
3-branes. Moreover, because the brane with
 negative brane tension may be an anti-gravity world~\cite{TSMS}, we assume that
 the observable brane which includes our world has positive
 tension.  
By the way, there are positive energy objects, namely D-branes and NS-branes,
that are well understood  and on which the gauge fields and matter fields can be 
localized so that the Standard Model fields can be placed there.

In this paper, we construct the general models with  parallel  3-branes and 
with the constant/sectional constant cosmology constant along the fifth dimension
on space-time $M^4\times R^1$, $M^4\times R^1/Z_2$, $M^4\times S^1$, and $M^4\times 
S^1/Z_2$. 
In general, if the space-time is compact, the sum of the brane tensions is zero,
if the space-time is non-compact, the sum of the brane tensions is positive if one
requires that the four-dimensional Planck scale is finite.
In addition, if the cosmology constant is constant along the fifth dimension, we
can define the index of the model: the sum of the brane tensions divided by
$6 k M_X^3$, where $k$ is defined in the following section. If the
space-time is $M^4\times R^1$, the index is $+1$, 
if space-time is $M^4\times R^1/Z_2$, the index is $+{1\over 2}$, and
if the fifth dimension is compact, the index is $0$. 
Moreover, for any point in 
$M^4 \times R^1$, $M^4 \times R^1/Z_2$, $M^4 \times S^1$ and $M^4 \times S^1/Z_2$,
 which is not belong to any brane and the section where the cosmology constant
 is zero, 
there is a neighborhood which is diffeomorphic to ( or a slice of ) $AdS_5$ space.
It seems to us the gauge hierarchy problem can be solved easily in these scenarios.
Assume we had $l+m + 1$ brane with position $y_{-l} < y_{-l+1} <...<
y_{m-1} < y_m$,  $\sigma(y)$, which is defined in the following sections, will
have minimal value at the brane with positive tension. Without loss of generality,
we assume $\sigma (y_j)$ is minimal. If $\sigma( y_i ) - \sigma(y_j)$ is not small
for all $i\neq j$, then, $M_{pl}^2/M_X^3$ will be proportional to $e^{-2\sigma(y_j)}$.
And if the observable brane is at $y= y_{i_o}$, then, the 4-dimensional GUT scale
in our world will be proportional to $M_{pl} \times e^{-(\sigma(y_{i_o}) -\sigma(y_j))}$
under some conditions, if $ \sigma(y_{i_o}) -\sigma(y_j) $ = 34.5 and 30, we will
push the four-dimensional GUT scale in our world to the TeV and $10^5$ GeV range, 
respectively. 
The 4-dimensional GUT scale on $j-th$ brane is the  highest, but, it does not
mean the $j-th$ brane has larger brane tension. Furthermore, if $\sigma (y_j) > 0$ and
the $j-th$ brane is the observable brane, we can also solve the gauge hierarchy problem,
but, the 5-dimensional Planck scale will be very large. In short,
the 5-dimensional GUT scale on each brane
can be indentified as the 5-dimensional Planck scale, but,
the 4-dimensional Planck scale is generated from the low 4-dimensional
GUT scale exponentially in our world.
 
We also give some simple models to show how to solve the gauge hierarchy problem.

This paper is organized as this way, in section 2, we discuss the
general models with constant cosmology constant along the fifth dimension on the
space-time $M^4 \times R^1$, $M^4 \times R^1/Z_2$,
$M^4 \times S^1$ and $M^4 \times S^1/Z_2$. In section 3,
we discuss the
general models with sectional constant cosmology constant along the fifth dimension
on the space-time $M^4 \times R^1$, $M^4 \times R^1/Z_2$,
$M^4 \times S^1$ and $M^4 \times S^1/Z_2$. The conclusion is
given in section 4. 

\section{General Models with Constant Cosmology Constant along the Fifth Dimension}

In this section, we would like to consider the cosmology constant is constant along the 
fifth dimension.
The models on the space-time $M^4 \times R^1$, $M^4 \times R^1/Z_2$,
$M^4 \times S^1$ and $M^4 \times S^1/Z_2$ will be discussed in the subsections 2.1, 2.2,
2.3, 2.4, respectively. By the way, generally speaking,
for any point in 
$M^4 \times R^1$, $M^4 \times R^1/Z_2$, $M^4 \times S^1$ and $M^4 \times S^1/Z_2$,
 which is not belong to any brane,
there is a neighborhood which is diffeomorphic to ( or a slice of ) $AdS_5$ space.
And $k > 0$ has been assumed in this section where $k$ is defined in the following 
subsections.

\subsection{Models on the Space-Time $M^4 \times R^1$}
In this subsection, we would like to consider the fifth dimension is $R^1$. Assume we have 
$m$ parallel 3-branes,
and their fifth coordinates are: $-\infty <  y_{1} < ... < y_{m-1} < y_{m} < +\infty$. 
The 5-dimensional metric in these  branes are:
\begin{equation}
\label{smmetric}
g_{\mu \nu}^{(i)} (x^{\mu}) \equiv G_{\mu \nu}(x^{\mu}, y=y_i) ~,~\,
\end{equation}
where $G_{AB}$ is the five-dimensional metric, and $A, B = \mu, y$~\footnote{
We assume that $G_{ \mu 5} = 0 $ here. In addition, if the fifth dimesnion has
$Z_2$ symmetry, i. e.,  the Lagrangian is invariant under the 
transformation $ y \leftrightarrow -y$, then, $G_{ \mu 5} = 0 $.}.

The classical Lagrangian is given by:

\begin{eqnarray}
S &=& S_{gravity} + S_B 
~,~\,
\end{eqnarray}
\begin{eqnarray}
S_{gravity} &=& 
\int d^4 x  ~dy~ \sqrt{-G} \{- \Lambda  + 
{1\over 2} M_X^3 R \} 
~,~\,
\end{eqnarray}
\begin{eqnarray}
S_B &=& \sum_{i=1}^{m} \int d^4 x \sqrt{-g^{(i)}} \{ {\cal L}_{i} 
-  V_{i} \} 
~,~\,
\end{eqnarray}
where $M_X$ is the 5-dimensional Planck scale,
$\Lambda $ is the cosmology constant, and 
$V_i$ where $i=1, ..., m$ is the brane tension.

The 5-dimensional Einstein equation for the  above action is:
\begin{eqnarray} 
\sqrt{-G} \left( R_{AB}-{1 \over 2 } G_{AB} R \right) &=& - \frac{1}{ M_X^3} 
[ \Lambda  \sqrt{-G} ~G_{AB} +  
\nonumber\\&& \sum_{i=1}^{m}
V_{i} \sqrt{-g^{(i)}} ~g_{\mu \nu}^{(i)} 
~\delta^\mu_M \delta^\nu_N ~\delta(y-y_i)  ] ~.~ \,
\end{eqnarray}
Assuming that there exists a solution that
 respects   4-dimensional 
Poincare invariance in the $x^{\mu}$-directions, one obtains
the 5-dimensional metric:
\begin{eqnarray} 
ds^2 = e^{- 2 \sigma(y)} \eta_{\mu \nu} dx^{\mu} dx^{\nu}
 + dy^2 ~.~\, 
\end{eqnarray}
With this metric, the Einstein equation reduces to:
\begin{eqnarray}
\sigma^{\prime 2} = { - {\Lambda } \over { 6 M_X^3}} 
 ~,~\, 
\end{eqnarray}
\begin{eqnarray}
 \sigma^{\prime \prime} =  \sum_{i=1}^{m}
{{V_{i}} \over\displaystyle {3 M_X^3 } } \delta (y-y_i)
~.~\, 
\end{eqnarray}

Now, we consider the general solutions. For $m$ is odd, assuming that
$m=2n+1$ where $n \geq 0$, we obtain two solutions $A$ and $B$:
\begin{equation}
\sigma (y)_A = \sum_{i=1}^{2n+1} (-1)^{i+1} k |y-y_i| + c 
~,~\,
\end{equation}
\begin{equation}
\sigma (y)_B = \sum_{i=1}^{2n+1} (-1)^{i} k |y-y_i| + c 
~.~\,
\end{equation}
For $m$ is even, assuming that
$m=2n$ where $n \geq 1$, we also obtain two solutions $C$ and $D$:
\begin{equation}
\sigma (y)_C = \sum_{i=1}^{2n} (-1)^{i+1} k |y-y_i| -k y + c 
~,~\,
\end{equation}
\begin{equation}
\sigma (y)_D = \sum_{i=1}^{2n} (-1)^{i} k |y-y_i| + k y + c 
~.~\,
\end{equation}
And one can easily obtain the cosmology constant:
\begin{equation}
\Lambda_i= -6 k^2 M_X^3 
~.~\,
\end{equation}
For solutions $A$ and $C$,  the bane tensions are:
\begin{equation}
V_i=  (-1)^{i+1} 6 k M_X^3 
~,~\,
\end{equation}
and for solutions $B$ and $D$, the brane tensions are:
\begin{equation}
V_i=  (-1)^{i} 6 k M_X^3 
~.~\,
\end{equation}

For the models with constant cosmology constant along the fifth dimension,
we can define the index:
the sum of brane tensions divided by $6 k M_X^3$. Therefore, solution $A$ has index $+1$,
solution $B$ has index $-1$, the solution $C$ and $D$ have index
$0$. If one required that the 4-dimensional Planck scale is finite,
only solution $A$ is reasonable. Therefore, we obtain the 
five-dimensional metric:
\begin{equation}
ds^2 = e^{-2 \sum_{i=1}^{2n+1} (-1)^{i+1} k |y-y_i| -2 c } 
\eta_{\mu \nu} dx^{\mu} dx^{\nu} + dy^2
~.~ \,
\end{equation}

And the corresponding four-dimensional Planck scale  is:
\begin{equation}
M_{pl}^2 = {1\over k}  M_X^3 \left( \sum_{i=1}^{2n+1} (-1)^{i+1}
e^{-2 \sigma (y_i)} \right) 
 ~.~ \,
\end{equation}

In addition, the four-dimensional GUT scale on i-th brane $M_{GUT}^{(i)}$ is
related to the five-dimensional GUT scale on i-th brane $M5_{GUT}^{(i)}$:
\begin{equation}
M_{GUT}^{(i)} = M5_{GUT}^{(i)} e^{-\sigma (y_i)} 
 ~.~ \,
\end{equation}
Our general assumption is that $ M5_{GUT}^{(i)} \equiv M_X $, for  
$i =1, ..., 2n+1$.

These models can be generalized to the models 
with $Z_2$ symmetry: 
one just requires that, $y_{n+1} =0$, $y_i= -y_{2n+2-i}$ for $1 \leq i \leq n$.

Now, we would like to give two simple models.

(I) Three brane model: the branes' positions are $y_1$, $y_2$, $y_3$, respectively, and  
the values of the brane 
tensions divided by $6 M_X ^3$ are: $ k$, $-k$, $k$, respectively. Therefore, we obtain:
\begin{equation}
\sigma (y) = \sum_{i=1}^{3} (-1)^{i+1} k |y-y_i| + c 
~,~\,
\end{equation}
\begin{equation}
\sigma (y_1) = k ( y_3- y_2 )+ c ~,~ \sigma (y_2) = k ( y_3- y_1 )+ c
~,~\,
\end{equation}
\begin{equation}
\sigma (y_3) = k ( y_2- y_1 )+ c 
~.~\,
\end{equation}

Without loss of generality, we assume $y_2-y_1 < y_3 -y_2$. The four-dimensional Planck 
scale is
\begin{equation}
M_{pl}^2 = {{M_X^3} \over k} e^{-2 \sigma(y_3) } \left( e^{-2 (\sigma(y_1)-\sigma(y_3))} - 
e^{-2 (\sigma(y_2)-\sigma(y_3))} + 1 \right)
~.~\,
\end{equation}

If  the brane with position $y_1$  is the observable brane which includes
our world, we can solve the gauge hierarchy problem. Assuming that  $e^{-2 
(\sigma(y_1)-\sigma(y_3))} < < 1$,
and $M_X=k$, we obtain:
\begin{equation}
M_{GUT}^{(1)} = M_{pl} e^{-k (y_3+y_1 -2 y_2)} 
 ~.~ \,
\end{equation}
So, we can push the GUT scale in our world to TeV scale and
$10^5$ GeV scale range if  $k (y_3+y_1 -2 y_2) $ = 34.5 and 30, 
respectively. And the value of $\sigma(y_3)=k ( y_2- y_1 )+ c$ determines the
relation between $M_{pl}$ and $M_X$.

If the brane with position $y_3$  is the observable brane which includes
our world, we can solve the gauge hierarchy problem, too. Assuming that
$M_{pl} = k $ and  $e^{-2 (\sigma(y_1)-\sigma(y_3))} < < 1$, we obtain:
\begin{equation}
M_{GUT}^{(3)} = M_{pl} e^{-{1\over 3}k ( y_2- y_1 )- {1\over 3} c  } 
 ~,~ \,
\end{equation}
with $k ( y_2- y_1 )+ c $ = 103.5 and 90, we can have the GUT scale in
our world at TeV scale and $10^5$ GeV scale, respectively.
Obviously, in this case, because of the exponential factor,
 the five dimensional Planck scale is very high,
$10^{48}$ GeV and $10^{44}$ GeV, respectively.

(II) Five brane model with $Z_2$ symmetry: the branes' positions are: $y_1=-y_5$, 
$y_2=-y_4$, $y_3=0$, $y_4$, $y_5$,
 respectively, and  the values of the brane 
tensions divided by $6 M_X ^3$ are: $ k$, $-k$, $k$,  $-k$, $k$, respectively. Therefore, 
we obtain:
\begin{equation}
\sigma (y) = k |y| +\sum_{i=4}^{5} (-1)^{i+1}  k ( |y-y_i|+ |y+y_i| ) + c 
~,~\,
\end{equation}
\begin{equation}
\sigma ( y_1)=\sigma ( y_5) = k y_5 + c ~,~\,
\end{equation}
\begin{equation}
\sigma ( y_2)=\sigma (y_4) = 2 k y_5 - k y_4 + c
~,~\,
\end{equation}
\begin{equation}
\sigma (y_3) = 2 k ( y_5- y_4 )+ c 
~.~\,
\end{equation}

Without loss of generality, we assume that, $y_5 > 2 y_4$. The four-dimensional Planck 
scale is:
\begin{equation}
M_{pl}^2 = {{M_X^3}\over k} e^{-2 k y_5 -2c }
\left(2- 2e^{-2 k (y_5 -y_4) } + e^{-2 k ( y_5 -2 y_4)} \right)
~.~\,
\end{equation}

If  the brane with position $y_3$  is the observable brane,
the gauge hierarchy problem can be solved. Assuming that  $ e^{-2 k ( y_5 -2 y_4)}  < < 1$,
and $M_X=k$, we obtain:
\begin{equation}
M_{GUT}^{(3)} = M_{pl} e^{-k (y_5 -2 y_4)} 
 ~.~ \,
\end{equation}
So, the GUT scale in our world will be pushed to TeV scale and
$10^5$ GeV scale range if  $k (y_5 -2 y_4) $ = 34.5 and 30, 
respectively. And the value of $k y_5 + c $ determines the
relation between $M_{pl}$ and $M_X$.

If the brane with position $y_1$  is the observable brane,
one can also solve the gauge hierarchy problem. Assuming that
$M_{pl} = k $ and  $ e^{-2 k ( y_5 -2 y_4)} < < 1$, one obtains:
\begin{equation}
M_{GUT}^{(1)} = M_{pl} e^{-{1\over 3}k y_5- {1\over 3} c  } 
 ~,~ \,
\end{equation}
with $k y_5 + c $ = 103.5 and 90, one obtains that the GUT scale in
our world is at TeV scale and $10^5$ GeV scale, respectively.
The five-dimensional Planck scale is
$10^{48}$ GeV and $10^{44}$ GeV, respectively.

\subsection{Models on the Space-Time $M^4 \times R^1/Z_2$}
Now, we consider the models on the space-time $M^4 \times R^1/Z_2$. 
The models with $Z_2$ symmetry in the subsection 2.1 
 can be generalized to the models in which the fifth dimension is $R^1/Z_2$.
If the models in last subsection has $Z_2$ symmetry, we just introduce  
the equivalence classes: $ y \sim - y$
 and $i-th ~brane \sim  (2n+2-i)-th ~brane$. 
After moduling the equivalence classes, we obtain the models on $M^4 \times R^1/Z_2$.
The trick  point is the tension of the brane at origin.  
We would like to explain the technical detail as this way:
the brane at origin which is $(n+1)$-th brane with tension
$ (-1)^{n} 6 k M_X^3$, we split it into two branes with equal tension 
$ (-1)^{n} 3 k M_X^3$, and with the positions $y=- \epsilon$ and
$ y = + \epsilon$, respectively. In the limit, $\epsilon \rightarrow 0$, we obtain the
original case. Therefore, ${{d\sigma(y)} \over\displaystyle 
dy}$ is zero at $ y=0$. And after moduling the equivalence classes, we obtain
the brane tension $V_{n+1}$ is half of the original value, i. e., 
$V_{n+1}=  (-1)^{n} 3 k M_X^3$. In general, using this method, 
we obtain ${{d\sigma(y)} \over\displaystyle dy}$ is zero at boundary. By the way,
 the index of this kind of the models is ${1\over 2}$.

For $n$ is odd, we obtain:
\begin{equation}
\sigma (y) = \sum_{i=n+2}^{2n+1} (-1)^{i+1} k |y-y_i| + c^{\prime} 
~,~\,
\end{equation} 
and for $n$ is even, we obtain:
\begin{equation}
\sigma (y) = k |y| + \sum_{i=n+2}^{2n+1} (-1)^{i+1} k |y-y_i| + c^{\prime} 
~,~\,
\end{equation} 
where 
\begin{equation}
c^{\prime} = c + \sum_{i=1}^{n} (-1)^{i+1} k y_{2n+2 -i}
~.~\,
\end{equation}
Because $c^{\prime}$ is just a constant, we will write it as
c in the following part. Similar convention will be used in the following subsections.

For the simplicity, we just renumber above subscript $i$ where $i=n+1, ..., 2n+1$
as $i-(n+1)$. So, we have $n+1$ parallel 3-branes, with
position $y_0=0 < y_1 <...<y_{n-1} < y_n < +\infty$.
 Then, $\sigma(y) $ can be written as:
\begin{equation}
\sigma (y) ={1\over 2} (1+ (-1)^n) k|y| + 
\sum_{i=1}^{n} (-1)^{i+n} k |y-y_i| + c 
~.~\,
\end{equation} 

One can easily obtain the cosmology constant
\begin{equation}
\Lambda = -6 k^2 M_X^3 
~,~\,
\end{equation}
and the bane tensions are:
\begin{equation}
V_0 = (-1)^n 3 k M_X^3 ~,~ V_i=  (-1)^{i+n} 6 k M_X^3 
~.~\,
\end{equation}

And the corresponding 4-dimensional Planck scale  is:
\begin{equation}
M_{pl}^2 = {1\over k}  M_X^3 \left( { {(-1)^n}\over 2} e^{-2 \sigma(y_0)} +
\sum_{i=1}^{n} (-1)^{i+n}
e^{-2 \sigma (y_i)} \right) 
 ~.~ \,
\end{equation}

In addition, the four-dimensional GUT scale on i-th brane $M_{GUT}^{(i)}$ is
related to the five-dimensional GUT scale on i-th brane $M5_{GUT}^{(i)}$:
\begin{equation}
M_{GUT}^{(i)} = M5_{GUT}^{(i)} e^{-\sigma (y_i)} \equiv M_X e^{-\sigma (y_i)}
 ~.~ \,
\end{equation}

Now, we discuss the example. Because the model with three branes is equivalent to 
the model with 
five 3-branes and $Z_2$ symmetry in the subsection 2.1, we donot rediscuss it here. 
We discuss the model with four 3-branes: the branes' positions are $y_0$, $y_1$, $y_2$, 
$y_3$, respectively, 
and  the values of the brane 
tensions divided by $6 M_X ^3$ are: $-{1\over 2} k$, $ k$, $-k$, $k$, respectively. 
Therefore, we obtain:
\begin{equation}
\sigma (y) = \sum_{i=1}^{3} (-1)^{i+1} k |y-y_i| + c 
~,~\,
\end{equation}
\begin{equation}
\sigma (y_0) = k ( y_3+ y_1 - y_2 )+ c ~,~ \sigma (y_1) = k ( y_3- y_2 )+ c
~,~\,
\end{equation}
\begin{equation}
\sigma (y_2) = k ( y_3 - y_1 )+ c ~,~ \sigma (y_3) = k ( y_2- y_1 )+ c
~.~\,
\end{equation}

Without loss of generality, we assume $y_2-y_1 > y_3 -y_2$. The four-dimensional Planck 
scale is
\begin{equation}
M_{pl}^2 = {{M_X^3}\over k} e^{-2 k( y_3 -y_2 )-2c  } \left(-{1\over 2} e^{-2k y_1} +1
- e^{-2k(y_2-y_1)} + e^{-2 k (2y_2-y_1-y_3)}  \right)
~.~\,
\end{equation}

If  the brane with position $y_3$  is the observable brane,
 we can solve the gauge hierarchy problem. Assuming that  $ e^{-2 k (2y_2-y_1-y_3)} < < 1$ 
and $M_X=k$, we obtain:
\begin{equation}
M_{GUT}^{(3)} = M_{pl} e^{-k (2y_2-y_1-y_3)} 
 ~.~ \,
\end{equation}
So, we can push the GUT scale in our world to TeV scale and
$10^5$ GeV scale range if  $ k (2y_2-y_1-y_3) $ = 34.5 and 30, 
respectively. And the value of $\sigma(y_1)= k ( y_3- y_2 )+ c$ determines the
relation between $M_{pl}$ and $M_X$.

If the brane with position $y_1$  is the observable brane,
 we can solve the gauge hierarchy problem, too. Assuming that
$M_{pl} = k $ and  $e^{-2 k (2y_2-y_1-y_3)} < < 1$, we obtain:
\begin{equation}
M_{GUT}^{(1)} = M_{pl} e^{-{1\over 3}k (y_3-y_2 )- {1\over 3} c  } 
 ~,~ \,
\end{equation}
with $k ( y_3- y_2 )+ c $ = 103.5 and 90, we can have the GUT scale in
our world at TeV scale and $10^5$ GeV scale, respectively.
The five dimensional Planck scale is 
$10^{48}$ GeV and $10^{44}$ GeV, respectively.

\subsection{Models on the Space-Time $M^4 \times S^1$}

In this subsection, we consider the models on the space-time $M^4 \times S^1$.
Because the adjacent two branes will have opposite brane tensions,
i. e., one has positive tension, and the other one has negative tension, we will have even
number of the branes and the index is $0$. Let us assume that we have
$2n$ branes, and their positions are $0 \leq y_1 < y_2 < ... < y_{2n-1}
< y_{2n} < 2 \pi \rho$, where $\rho$ is the radius of the fifth dimension. The Lagrangian, 
Einstein equation, and the
differential equations of $\sigma(y)$ are similar to those in 
subsection 2.1 except we must have even number of branes, $0 \leq
y \leq 2 \pi \rho$, and $\sigma(0)=\sigma(2 \pi \rho)$. 

If $y_1=0$, then, the two solutions $A$ and $B$ of $\sigma (y)$ are 
\begin{equation}
\sigma (y)_A = \sum_{i=2}^{2n} (-1)^{i+1} k |y-y_i| + c 
~,~\,
\end{equation}
\begin{equation}
\sigma (y)_B = \sum_{i=2}^{2n} (-1)^{i} k |y-y_i| + c 
~.~\,
\end{equation}
If $y_1 \neq 0$, then, the two solutions $C$ and $D$ are: 
\begin{equation}
\sigma (y)_C = \sum_{i=1}^{2n} (-1)^{i+1} k |y-y_i| - k y + c 
~,~\,
\end{equation}
\begin{equation}
\sigma (y)_D = \sum_{i=1}^{2n} (-1)^{i} k |y-y_i| + k y + c 
~.~\,
\end{equation}
And one can easily obtain the cosmology constant
\begin{equation}
\Lambda_i= -6 k^2 M_X^3 
~.~\,
\end{equation}
For solution $A$ and $C$,  the bane tensions are:
\begin{equation}
V_i=  (-1)^{i+1} 6 k M_X^3 
~,~\,
\end{equation}
and for solution $B$ and $D$, the brane tensions are:
\begin{equation}
V_i=  (-1)^{i} 6 k M_X^3 
~.~\,
\end{equation}

The constraint is $\sigma (0)= \sigma (2 \pi \rho)$. So, we
obtain the constraint equations on solutions $A, B, C, D$, respectively:
\begin{equation}
\sum_{i=2}^{2n} (-1)^{i+1}  y_i = -  \pi \rho 
~,~\,
\end{equation}
\begin{equation}
\sum_{i=2}^{2n} (-1)^{i}  y_i =   \pi \rho
~,~\,
\end{equation} 
\begin{equation}
\sum_{i=1}^{2n} (-1)^{i+1}  y_i = -  \pi \rho 
~,~\,
\end{equation} 
\begin{equation}
\sum_{i=1}^{2n} (-1)^{i}  y_i =    \pi \rho
~.~\,
\end{equation} 

The corresponding 4-dimensional Planck scale for solutions $A$ and $C$ is:
\begin{equation}
M_{pl}^2 = {1\over k}  M_X^3 \left( \sum_{i=1}^{2n} (-1)^{i+1}
e^{-2 \sigma(y_i)_F} \right) 
 ~,~ \,
\end{equation}
where $F= A$, or $C$.
And the corresponding 4-dimensional Planck scale for solutions $B$ and $D$ is:
\begin{equation}
M_{pl}^2 = {1\over k}  M_X^3 \left( \sum_{i=1}^{2n} (-1)^{i}
e^{-2 \sigma (y_i)_F} \right) 
 ~,~ \,
\end{equation}
where $F=B$ or $D$.

In addition, the four-dimensional GUT scale on i-th brane $M_{GUT}^{(i)}$ is
related to the five-dimensional GUT scale on i-th brane $M5_{GUT}^{(i)}$:
\begin{equation}
M_{GUT}^{(i)} = M5_{GUT}^{(i)} e^{-\sigma (y_i)_F} \equiv M_X e^{-\sigma (y_i)_F}
 ~,~ \,
\end{equation}
where $F=A, B, C, D$.

In addition, we can consider the models with $Z_2$ symmetry. The $Z_2$ symmetry is defined 
by the
transformation of $y \leftrightarrow 2 \pi \rho - y $, and we require that the metric and 
the Lagrangian are
invariant under this transformation. With $Z_2$ symmetry, we must have a brane at position 
$0$,
and a brane at position $\pi \rho$. Therefore, only solution (A) and (B) are possible. So,
$y_1 = 0$, $y_{n+1} = \pi \rho$, $y_i = 2 \pi \rho - y_{2n+2 -i}$ for $i=2,...,n$.

Let us give an explicit model.
We discuss four  3-brane model: the branes' positions are: $y_1=0$, $y_2$, $y_3$, $y_4$, 
respectively, and 
the values of the brane 
tensions divided by $6 M_X ^3$ are: $ k$, $ -k$, $k$, $-k$, respectively. Therefore, we 
obtain:
\begin{equation}
\sigma (y) = \sum_{i=2}^{4} (-1)^{i+1} k |y-y_i| + c 
~,~\,
\end{equation}
and the constraint equation is:
\begin{equation}
y_3= y_2 + y_4 - \pi \rho
~.~\,
\end{equation}
So, one can easily obtain:
\begin{equation}
\sigma (y_1) = -k \pi \rho + c ~,~ \sigma (y_2) = - k ( \pi \rho -y_2)+ c
~,~\,
\end{equation}
\begin{equation}
\sigma (y_3) = - k ( y_4-y_2 )+ c ~,~ \sigma (y_4) = - k ( y_4- \pi \rho )+ c
~.~\,
\end{equation}

Without loss of generality, we assume $y_4-y_2 < \pi \rho$. The four-dimensional Planck 
scale is
\begin{equation}
M_{pl}^2 = {{M_X^3}\over k} e^{2 \pi  k \rho -2c  } \left(1- e^{-2k y_2} + 
e^{-2 k ( \pi \rho - y_4 +y_2) } - e^{-2 k ( 2 \pi \rho - y_4)}
\right)
~.~\,
\end{equation}

If  the brane with position $y_3$  is the observable brane,
 we can solve the gauge hierarchy problem. Assuming that  $ e^{-2 k ( \pi \rho - y_4 +y_2) 
} < < 1$,
$e^{-2k y_2} < < 1 $,
and $M_X=k$, we obtain:
\begin{equation}
M_{GUT}^{(3)} = M_{pl} e^{- k ( \pi \rho - y_4 +y_2)} 
 ~.~ \,
\end{equation}
So, we can push the GUT scale in our world to TeV scale and
$10^5$ GeV scale range if  $k ( \pi \rho - y_4 +y_2) $ = 34.5 and 30, 
respectively. And the value of $ \pi \rho k - c$ determines the
relation between $M_{pl}$ and $M_X$.

\subsection{Models on the Space-Time $M^4 \times S^1/Z_2$}
The models  with $Z_2$ symmetry in the subsection 2.3 can be generalized to the models
on the space-time $M^4\times S^1/Z_2$.  
For the models with $Z_2$ symmetry in 2.3, we introduce the equivalence
classes: $y \sim 2 \pi \rho - y$, and $i-th ~brane \sim (2n+2-i)-th ~brane$. Moduling 
the equivalence classes, we obtain the models on $M^4 \times S^1/Z_2$ with $n+1$ brane. 
Using our splitting boundary branes method as before, we obtain the tensions of the 
boundary branes will be half of their original values. So, the index in this kind of the
models is zero, too,
which is the only constraint. In short, after moduling the equivalence 
classes, we will have $n+1$ branes with positions $y_0=0 < y_1 < ... < y_{n-1} < y_{n} =\pi 
\rho$.

The general solution of $\sigma (y)$ is  
\begin{equation}
\sigma (y) = sign(V_0) \left(\sum_{i=1}^{n-1} (-1)^{i} k |y-y_i|+ {1\over 2} (1+(-1)^{n+1}) 
k y \right) + c 
~,~\,
\end{equation}
where $sign(V_i) = {|V_i|/V_i}$, for $i=1, ..., n$.

The corresponding 4-dimensional Planck scale is:
\begin{equation}
M_{pl}^2 = {{sign(V_0)} \over k}  M_X^3 \left( {1 \over {2}} e^{-2 \sigma(y_0)} +
\sum_{i=1}^{n-1} (-1)^{i}
e^{-2 \sigma(y_i)} + {{(-1)^{n}}  \over {2}} e^{-2 \sigma(y_{n}) } \right) 
 ~.~ \,
\end{equation}
In addition, the four-dimensional GUT scale on i-th brane $M_{GUT}^{(i)}$ is
related to the five-dimensional GUT scale on i-th brane $M5_{GUT}^{(i)}$:
\begin{equation}
M_{GUT}^{(i)} = M5_{GUT}^{(i)} e^{-\sigma (y_i)} \equiv M_X e^{-\sigma (y_i)}
 ~.~ \,
\end{equation}

Here,  we will give two simple models. 

(I) Although the model with three branes is equivalent to the model with four 3-branes and 
with $Z_2$
symmetry in the subsection 2.3. Here, we also discuss it, because of its simplicity and
phenomenology interesting. 
Assuming we have three branes with
  positions: $y_0=0$, $y_1$, $y_2=\pi \rho$,  respectively, and  the values of the brane 
tensions divided by $6 M_X ^3$ are: ${1\over 2} k$, $ -k$,  ${1\over 2} k$, respectively,  
we obtain:
\begin{equation}
\sigma (y) = - k |y-y_1| + c 
~,~\,
\end{equation}
\begin{equation}
\sigma (y_0) = - k y_1+ c ~,~ \sigma (y_1) = c
~,~\,
\end{equation}
\begin{equation}
\sigma (y_2) = - k (\pi \rho - y_1) + c 
~.~\,
\end{equation}

Without loss of generality, we assume $\pi \rho  < 2 y_1$. The four-dimensional Planck 
scale is
\begin{equation}
M_{pl}^2 = {{M_X^3}\over {2k}} e^{2 k y_1-2 c  } \left( 1
-2 e^{-2 k y_1} + e^{-2 k( 2 y_1 - \pi \rho)}   \right)
~.~\,
\end{equation}

If  the brane with position $y_2$  is the observable brane,
 we can solve the gauge hierarchy problem. Assuming that  $ e^{-2 k( 2 y_1 - \rho)} << 1 $,
and $M_X= 2 k$, we obtain:
\begin{equation}
M_{GUT}^{(2)} = M_{pl} e^{-k (2 y_1 - \pi \rho)} 
 ~.~ \,
\end{equation}
So, we can push the GUT scale in our world to TeV scale and
$10^5$ GeV scale range if  $ k (2 y_1 - \pi \rho)$ = 34.5 and 30, 
respectively. And the value of $ k y_1- c  $ determines the
relation between $M_{pl}$ and $M_X$.

(II) Model with four  3-branes: the branes' positions are: $y_0=0$, $y_1$, $y_2$, $y_3=\pi 
\rho$, 
respectively, and  the values of the brane 
tensions divided by $6 M_X ^3$ are: ${1\over 2} k$, $ -k$, $k$, $-{1\over 2} k$, 
respectively. Therefore, we obtain:
\begin{equation}
\sigma (y) = k y + \sum_{i=1}^{2} (-1)^{i} k |y-y_i| + c 
~,~\,
\end{equation}
\begin{equation}
\sigma (y_0) = k ( y_2- y_1  )+ c ~,~ \sigma (y_1) = k   y_2 + c
~,~\,
\end{equation}
\begin{equation}
\sigma (y_2) = k  y_1+ c ~,~ \sigma (y_3) = k ( \pi \rho - y_2 + y_1 )+ c
~.~\,
\end{equation}

Without loss of generality, we assume $2 y_1 < y_2$. The four-dimensional Planck scale is
\begin{equation}
M_{pl}^2 = {{M_X^3}\over k} e^{-2 k y_1-2 c  } \left( {1\over 2} e^{-2k (y_2-2y_1)}
-e^{-2k(y_2-y_1)} +1-{1\over 2} e^{-2k (\pi \rho - y_2)}   \right)
~.~\,
\end{equation}

If  the brane with position $y_0$  is the observable brane,
 we can solve the gauge hierarchy problem. Assuming that  $ e^{-2k (y_2-2y_1)} < < 1, 
e^{-2k (\pi \rho - y_2)} << 1 $,
and $M_X=k$, we obtain:
\begin{equation}
M_{GUT}^{(0)} = M_{pl} e^{-k (y_2-2y_1)} 
 ~.~ \,
\end{equation}
So, the GUT scale in our world will be pushed to TeV scale and
$10^5$ GeV scale range if  $ k (y_2-2 y_1) $ = 34.5 and 30, 
respectively. And the value of $\sigma(y_2)= k y_1 + c$ determines the
relation between $M_{pl}$ and $M_X$.

If the brane with position $y_2$  is the observable brane,
one can also solve the gauge hierarchy problem. Assuming that
$M_{pl} = k $, $ e^{-2k (y_2-2y_1)} < < 1$, and $ e^{-2k (\pi \rho - y_2)} << 1 $ , one 
obtains:
\begin{equation}
M_{GUT}^{(2)} = M_{pl} e^{-{1\over 3} k y_1- {1\over 3} c  } 
 ~,~ \,
\end{equation}
with $k y_1 + c $ = 103.5 and 90, one will have the GUT scale in
our world at TeV scale and $10^5$ GeV scale range, respectively.
The five-dimensional Planck scale is $10^{48}$ GeV and $10^{44}$ GeV, respectively.

\section{General Models with Sectional Constant Cosmology Constant along the Fifth 
Dimension}
In this section, we consider the general models with sectional constant cosmology constant 
along the
fifth dimension.
The models on the space-time $M^4 \times R^1$, $M^4 \times R^1/Z_2$,
$M^4 \times S^1$ and $M^4 \times S^1/Z_2$ will be discussed in the subsection 3.1, 3.2,
3.3, 3.4, respectively. In addition, we assume that $\chi_{i, i+1} \neq 0$ in the 
discussions of the explicit models, where $\chi_{i, i+1}$ is defined in the following 
subsections.

\subsection{Models on the Space-Time $M^4\times R^1$}
First, we consider the fifth dimension is $R^1$. Assuming we have $l+m+1$ parallel 
3-branes, and
their fifth coordinastes are: $-\infty < y_{-l} < y_{-l+1} < ...< y_{-1} < y_0
< y_{1} < ... < y_{m-1} < y_{m} < +\infty$.  
The 5-dimensional metric in these  branes are:
\begin{equation}
\label{smmetric}
g_{\mu \nu}^{(i)} (x^{\mu}) \equiv G_{\mu \nu}(x^{\mu}, y=y_i) ~,~\,
\end{equation}
where $G_{AB}$ is the five-dimensional metric, and $A, B = \mu, y$.

The classical Lagrangian is given by:

\begin{eqnarray}
S &=& S_{gravity} + S_B 
~,~\,
\end{eqnarray}
\begin{eqnarray}
S_{gravity} &=& 
\int d^4 x  ~dy~ \sqrt{-G} \{- \Lambda (y) + 
{1\over 2} M_X^3 R \} 
~,~\,
\end{eqnarray}
\begin{eqnarray}
S_B &=& \sum_{i=-l}^{m} \int d^4 x \sqrt{-g^{(i)}} \{ {\cal L}_{i} 
-  V_{i} \} 
~,~\,
\end{eqnarray}
where $M_X$ is the 5-dimensional Planck scale,
$\Lambda (y)$ is the cosmology constant, and 
$V_i$ where $i=-l, ..., m$ is the brane tension.
The $\Lambda(y)$ is defined as the following:
\begin{eqnarray}
\Lambda (y) &=& \sum_{i=1}^m \Lambda_i \left(\theta (y-y_{i-1}) - \theta (y-y_i) \right)
+ \Lambda_{+\infty} \theta (y-y_m)
\nonumber\\&& +
\sum_{i=-l+1}^0 \Lambda_i \left(\theta (-y+y_i) - \theta (-y+y_{i-1}) \right)
+ \Lambda_{-\infty} \theta ( -y + y_{-l} ) 
~,~\,
\end{eqnarray}
where $\theta (x) = 1$ for $x \geq 0$ and $\theta (x) = 0$ for $x < 0$. From
advanced calculus, we know that $\Lambda(y) $ is piece-wise continuous or
sectionally continuous, exactly speaking, $\Lambda(y) $ is sectional
constant.

The 5-dimensional Einstein equation for the  above action is:
\begin{eqnarray} 
\sqrt{-G} \left( R_{AB}-{1 \over 2 } G_{AB} R \right) &=& - \frac{1}{ M_X^3} 
[ \Lambda (y) \sqrt{-G} ~G_{AB} +  
\nonumber\\&& \sum_{i=-l}^{m}
V_{i} \sqrt{-g^{(i)}} ~g_{\mu \nu}^{(i)} 
~\delta^\mu_M \delta^\nu_N ~\delta(y-y_i)  ] ~.~ \,
\end{eqnarray}
Assuming that there exists a solution that
 respects   4-dimensional 
Poincare invariance in the $x^{\mu}$-directions, one obtains
the 5-dimensional metric:
\begin{eqnarray} 
ds^2 = e^{- 2 \sigma(y)} \eta_{\mu \nu} dx^{\mu} dx^{\nu}
 + dy^2 ~.~\, 
\end{eqnarray}
With this metric, the Einstein equation reduces to:
\begin{eqnarray}
\sigma^{\prime 2} = { - {\Lambda (y) } \over { 6 M_X^3}} ~,~
 \sigma^{\prime \prime} =  \sum_{i=-l}^{m}
{{V_{i}} \over\displaystyle {3 M_X^3 } } \delta (y-y_i)
~.~\, 
\end{eqnarray}
So, $\sigma^{\prime }$ is also sectionally continuous or
sectional constant.

The general solution to the above differential equations is:
\begin{equation}
\sigma (y) = \sum_{i=-l}^m k_i |y-y_i| + k_c y + c 
~,~\,
\end{equation}
where $k_c$ and c are constants, and $k_i \ne 0$ for $i=-l, ..., m$.
The relations between the $k_i$ and $V_i$, 
and the relations between the $k_i$  and $\Lambda_i$ are:
\begin{equation}
V_i= 6 k_i M_X^3
~,~\,
\end{equation}
\begin{equation}
\Lambda_i= -6 M_X^3 (\sum_{j=i}^m k_j - \sum_{j=-l}^{i-1} k_j-k_c)^2
~,~\,
\end{equation}
\begin{equation}
\Lambda_{-\infty}= -6 M_X^3 (\sum_{j=-l}^m k_j-k_c)^2
~,~ \Lambda_{+\infty}= -6 M_X^3 (\sum_{j=-l}^m k_j+ k_c)^2
~.~\,
\end{equation}
Therefore, the cosmology constant is negative except the section(s) 
between the two branes where the cosmology 
constant is zero, then, for any point in 
$M^4 \times R^1$, 
 which is not belong to any brane and that kinds of sections, 
there is a neighborhood which is diffeomorphic to ( or a slice of ) $AdS_5$ space.
Similarly,  this statement is correct for the models on the space-time $M^4 \times 
R^1/Z_2$,
$M^4 \times S^1$ and $M^4 \times S^1/Z_2$.  Moreover, the
cosmology constant and brane tensions should satisfy above equations. In order to obtain 
finite
4-dimensional Planck scale, we obtain the constraint: $ \sum_{j=-l}^m k_j > | k_c |$.
So, the sum of the brane tensions is positive.

The general bulk metric is:
\begin{equation}
ds^2 = e^{-2 \sum_{i=-l}^m k_i |y-y_i| -2 k_c y - 2 c} \eta_{\mu \nu} dx^{\mu} dx^{\nu} + 
dy^2
~.~ \,
\end{equation}

And the corresponding 4-dimensional Planck scale  is:
\begin{equation}
M_{pl}^2 =  M_X^3 \left( T_{-\infty, -l} + T_{m, +\infty} + \sum_{i=-l}^{m-1} T_{i, i+1} 
\right) 
 ~,~ \,
\end{equation}
where 
\begin{equation}
T_{-\infty, -l} = {1 \over\displaystyle {2 \chi_{-\infty}}} e^{-2 \sigma (y_{-l})}
 ~,~ 
T_{m, +\infty} = {1 \over\displaystyle {2 \chi_{+\infty}}} e^{-2 \sigma (y_{m})}
 ~,~ \,
\end{equation}
if $\chi_{i, i+1} \neq 0$, then
\begin{equation}
T_{i, i+1} = {1 \over\displaystyle {2 \chi_{i, i+1}}} \left( e^{-2 \sigma (y_{i+1})}
 - e^{-2 \sigma (y_i)} \right)
 ~,~ \,
\end{equation}
and if $\chi_{i, i+1} =0$, then
\begin{equation}
T_{i, i+1} = (y_{i+1}- y_i) e^{-2  \sigma (y_i)}
 ~,~ \,
\end{equation}
where
\begin{equation}
\chi_{-\infty} = \sum_{j=-l}^m k_j - k_c ~,~ 
\chi_{+\infty} = \sum_{j=-l}^m k_j + k_c
 ~,~ \,
\end{equation}
\begin{equation}
\chi_{i, i+1} =
\sum_{j=i+1}^m k_j - \sum_{j=-l}^i k_j - k_c
 ~.~ \,
\end{equation}
By the way, one can easily prove that $T_{i, i+1}$ is positive, which makes sure that
the 4-dimensional Planck scale is positive.

In addition, the four-dimensional GUT scale on i-th brane $M_{GUT}^{(i)}$ is
related to the five-dimensional GUT scale on i-th brane $M5_{GUT}^{(i)}$:
\begin{equation}
M_{GUT}^{(i)} = M5_{GUT}^{(i)} e^{-\sigma (y_i)} 
 ~.~ \,
\end{equation}
In this paper, we assume that $ M5_{GUT}^{(i)} \equiv M_X $, for  $i =-l, ..., m$.

Now, in general,  we would like to explain how to solve the gauge hierarchy problem.
Using fact that $T_{-\infty, -l} >0,  T_{m, +\infty}  > 0,   T_{i, i+1} > 0$ where
$i=-l, ..., m-1$, and
noticing that $ \sum_{j=-l}^m k_j > | k_c |$, we can prove that if $ \sigma (y_j) $
is the minimal value of $ \sigma(y_i)$ for all $i$, then the $j-th$ brane will
have positive brane tension. If $e^{-2(\sigma( y_i ) - \sigma(y_j))} < < 1$
for all $i\neq j$, then,  we obtain: 
\begin{equation}
M_{pl}^2 = {{M_X^3}\over {m_c}} e^{-2 \sigma(y_j)}  
~,~\,
\end{equation}
where $m_c$ is a function of  $k_c$ and $k_i$ for all $i$.
And assuming the observable brane is at $y= y_{i_o}$ and $M_X =m_c$, we obtain: 
\begin{equation}
M_{GUT}^{(i_o)} = M_{pl} e^{- (\sigma(y_{i_o}) -  \sigma(y_j))} 
 ~.~ \,
\end{equation}
So, if $ \sigma(y_{i_o}) -\sigma(y_j) $ = 34.5 and 30, we will
push the four-dimensional GUT scale in our world to the TeV and $10^5$ GeV range, 
respectively. 
Of course, the 4-dimensional GUT scale on $j-th$ brane is the  highest, but, the $j-th$ 
brane does 
not need to have larger brane tension. Furthermore, if $\sigma (y_j) > 0$,   and if
the $j-th$ brane is the observable brane, we can also solve the gauge hierarchy problem.
Assume that $M_{pl}= m_c$, we obtain:
\begin{equation}
M_{GUT}^{(j)} = M_{pl} e^{-{1\over 3} \sigma(y_j)  } 
 ~,~ \,
\end{equation}
with $\sigma(y_j)$ = 103.5 and 90, we can have the GUT scale in
our world at TeV scale and $10^5$ GeV scale, respectively.
Obviously, in this case, because of the exponential factor,
 the five-dimensional Planck scale is very high,
$10^{48}$ GeV and $10^{44}$ GeV, respectively. 

These models can be generalized to the models 
with $Z_2$ symmetry. Because of $Z_2$ symmetry, $k_c=0$.  There are two kinds of such 
models, one is the odd number
of the branes, the other is the even number of the branes. For the first one,
one just requires that $k_{-i} = k_i$, $ y_{-i} = - y_i$,
 and $m=l$. For the second case,
one just requires that $k_{-i} = k_i$, $y_{-i} = -y_i$,
 $m=l$, and $k_0=0$ (no number 0 brane). 
 
Now, we discuss three models: 
 
(I) First, we consider the model with only one brane whose position is $y_0$ and  whose 
value of the brane 
tension divided by $6 M_X^3$ is  $k_0$. Therefore, we obtain:
\begin{equation}
\sigma (y) = k_0 |y-y_0| + k_c y + c 
~.~\,
\end{equation}
And the constraint is $ k_0 > |k_c| $. One can easily obtain:
\begin{equation}
\sigma (y_0) = k_c y_0+ c
~.~\,
\end{equation}

The four-dimensional Planck scale is
\begin{equation}
M_{pl}^2 = {M_X^3} {{k_0}\over\displaystyle {k_0^2-k_c^2}} e^{-2 k_c y_0 - 2c}
~.~\,
\end{equation}

Assuming that
$M_{pl} = {{k_0^2-k_c^2} \over\displaystyle {k_0}} $, we obtain:
\begin{equation}
M_{GUT}^{(0)} = M_{pl} e^{-{1\over 3} k_c y_0- {1\over 3} c  } 
 ~,~ \,
\end{equation}
with $k_c y_0 + c $ = 103.5 and 90, the GUT scale in
our world will be pushed to the TeV scale and $10^5$ GeV scale range, respectively.
The five-dimensional Planck scale is 
$10^{48}$ GeV and $10^{44}$ GeV, respectively. By the way, if $k_c$ = 0, 
the constant $c$ is the key factor, which is often thought physical irrelevant.
The model with $k_c=c=0$ was discussed in [3] .

In addition,  we can obtain the exact relation between $M_{pl}$ and $ M_{GUT}^{(0)}$:
\begin{equation}
M_{pl} = M_{GUT}^{(0)} \sqrt {{M_X k_0} \over\displaystyle {k_0^2 -k_c^2}}
 ~.~ \,
\end{equation}
In order to solve the gauge hierarchy problem, in general, we might require that $M_X$ is 
very large compare to 
$k_0$ and $k_c$, or $k_0^2 -k_c^2$ is very smalll.

(II) The model with two 3-branes: the branes' positions are $y_0$, $y_1$, respectively, and 
 the values of the brane 
tensions divided by $6 M_X ^3$ are: $ k_0 $, $k_1$, respectively. Therefore, we obtain:
\begin{equation}
\sigma (y) = k_0 | y-y_0 | + k_1 | y- y_1| + k_c y + c 
~,~\,
\end{equation}
The constraint is $ k_0 + k_1 > |k_c|$. One can easily obtain:  
\begin{equation}
\sigma (y_0) = k_1 ( y_1 -y_0 ) + k_c y_0 + c
~,~\,
\end{equation}
\begin{equation}
\sigma (y_1) = k_0 ( y_1 - y_0) + k_c y_1 + c
~.~\,
\end{equation}

The four-dimensional Planck scale is
\begin{equation}
M_{pl}^2 = {M_X^3}  \left({{ k_0} \over\displaystyle { k_0^2 - (k_1-k_c)^2}} e^{-2 
\sigma(y_0)} + 
{{ k_1} \over\displaystyle { k_1^2 - (k_0+ k_c)^2}} e^{-2 \sigma(y_1)} \right)
~.~\,
\end{equation}

If $k_0 > 0$, $k_1 > 0$, without loss of generality,assuming that $ \sigma(y_0) < 
\sigma(y_1) $
and $e^{-2 (\sigma(y_1)-\sigma(y_0))} << 1$, we obtain
\begin{equation}
M_{pl}^2 = {M_X^3} {{ k_0} \over\displaystyle { k_0^2 - (k_1-k_c)^2}} e^{-2 \sigma(y_0)}
~.~\,
\end{equation}
If the brane with position $y_1$ is the observable brane, assuming
 $M_X = { { k_0^2 - (k_1-k_c)^2} \over\displaystyle {k_0}}$, we obtain:
\begin{equation}
M_{GUT}^{(1)} = M_{pl} e^{-(\sigma(y_1) - \sigma(y_0)) } 
 ~.~ \,
\end{equation}
So, we can push the GUT scale in our world to TeV scale and
$10^5$ GeV scale range if  $(k_0 -k_1+k_c) ( y_1 -y_0 )$ = 34.5 and 30, 
respectively. And the value of $\sigma(y_0)$ determines the
relation between $M_{pl}$ and $M_X$.
Explicit example: $k_0=k_1=k_c > 0$, $ k_c( y_1 -y_0 )$ = 34.5 and 30.
From this explicit example, we conclude that
$k_c$ is also an important factor to solve the gauge hierarchy problem.

If the brane with position $y_0$  is the observable brane,
 we can solve the gauge hierarchy problem only when $ \sigma(y_0) > 0$. Assuming that
$M_{pl} = { { k_0^2 - (k_1-k_c)^2} \over\displaystyle {k_0}} $ and  
$e^{-2 (\sigma(y_1)-\sigma(y_0))} < < 1$, we obtain:
\begin{equation}
M_{GUT}^{(0)} = M_{pl} e^{-{1\over 3} \sigma(y_0)  } 
 ~,~ \,
\end{equation}
with $\sigma(y_0)$ = 103.5 and 90, we can have the GUT scale in
our world at TeV scale and $10^5$ GeV scale, respectively.
The five-dimensional Planck scale is
$10^{48}$ GeV and $10^{44}$ GeV, respectively. 

If $k_0 > 0, k_1 < 0$, the solution to the gauge hierarchy problem is similar to
that in just above paragraph.

(III) Model with three 3-branes: the branes'
 positions are $y_1$, $y_2$, $y_3$, respectively, and  the values of the brane 
tensions divided by $6 M_X ^3$ are: $ k_1$, $k_2$, $k_3$, respectively. And the constraint
is $ k_1 + k_2 + k_3 > |k_c|$. Therefore, we obtain:
\begin{equation}
\sigma (y) = \sum_{i=1}^{3}  k_i |y-y_i| + k_c y + c 
~,~\,
\end{equation}
\begin{equation}
\sigma (y_1) = k_2 ( y_2- y_1 )+k_3 ( y_3 - y_1) + k_c y_1 + c 
~,~\,
\end{equation}
\begin{equation}
\sigma (y_2) = k_1 ( y_2- y_1 )+k_3 ( y_3 - y_2) + k_c y_2 + c 
~,~\,
\end{equation}
\begin{equation}
\sigma (y_3) = k_1 ( y_3- y_1 )+k_2 ( y_3 - y_2) + k_c y_3 + c 
~.~\,
\end{equation}

The four-dimensional Planck scale is
\begin{eqnarray}
M_{pl}^2 &=& {{M_X^3} \over 2}  \left( {1\over\displaystyle {k_1 + k_2 + k_3 -k_c}} 
e^{-2\sigma(y_1)}
+ {1\over\displaystyle { k_2 + k_3 -k_c-k_1}} (  e^{-2\sigma(y_2)}-e^{-2\sigma(y_1)})
 \right.\nonumber\\&&\left.
+ {1\over\displaystyle {  k_3 -k_c-k_1-k_2}} (  e^{-2\sigma(y_3)}-e^{-2\sigma(y_2)})
 \right.\nonumber\\&&\left.
+ {1\over\displaystyle {k_1 + k_2 + k_3 + k_c}} e^{-2\sigma(y_3)}
 \right)
~.~\,
\end{eqnarray}

If there exist at least two branes which have positive tensions. Without loss of 
generality, 
we assume the brane
with position $y_1$ has positive tension, $ \sigma(y_1) <  \sigma(y_2)$,
and $ \sigma(y_1) <  \sigma(y_3)$. If $ e^{-2(\sigma(y_2) -  \sigma(y_1))} << 1$
and $ e^{-2(\sigma(y_3) -  \sigma(y_1))} << 1$, we obtain
\begin{equation}
M_{pl}^2 = {M_X^3} {{k_1}\over\displaystyle {k_1^2- (k_2 + k_3 -k_c)^2}} e^{-2\sigma(y_1)}
~.~\,
\end{equation}
If  the brane with position $y_j$ where $j=2$ or $3$ is the observable brane, and 
assuming $M_X = {{k_1^2- (k_2 + k_3 -k_c)^2} \over\displaystyle {k_1}} $, we obtain:
\begin{equation}
M_{GUT}^{(j)} = M_{pl} e^{- (\sigma(y_j) -  \sigma(y_1))} 
 ~.~ \,
\end{equation}
So, the GUT scale in our world will be at TeV scale and
$10^5$ GeV scale range if  $ \sigma(y_j) -  \sigma(y_1) $ = 34.5 and 30, 
respectively. And the value of $\sigma(y_1)$ determines the
relation between $M_{pl}$ and $M_X$. 
For an explicit example, $y_1= -y_3$, $y_2 =0 $, and $k_1 = k_3 = k_c$, $ k_2 = -{1\over 2} 
k_c$.
\begin{equation}
\sigma (y_1) = -{1\over 2} k_c y_3 + c 
~,~
\sigma (y_2) = 2 k_c y_3 + c 
~,~\,
\end{equation}
\begin{equation}
\sigma (y_3) = {5\over 2} k_c y_3 + c 
~.~\,
\end{equation}
If $M_X = {3\over 4} k_c$ and the brane with position $y_3$ is the observable brane, we 
obtain 
\begin{equation}
M_{GUT}^{(3)} = M_{pl} e^{- 3 k_c y_3} 
 ~.~ \,
\end{equation}
So, one can easily solve the gauge hierarchy problem.

If the brane with position $y_1$  is the observable brane,
 we can solve the gauge hierarchy problem only if $\sigma(y_1) >0 $. Assuming that
$M_{pl} = {{k_1^2- (k_2 + k_3 -k_c)^2} \over\displaystyle {k_1}} $ and  $e^{-2 
(\sigma(y_j)-\sigma(y_1))} < < 1$
where $j=2, 3$, we obtain:
\begin{equation}
M_{GUT}^{(1)} = M_{pl} e^{-{1\over 3} \sigma(y_1)  } 
 ~,~ \,
\end{equation}
if $\sigma (y_1) $ = 103.5 and 90, the GUT scale in
our world will be about TeV and $10^5$ GeV, respectively.
The five-dimensional Planck scale is 
$10^{48}$ GeV and $10^{44}$ GeV, respectively. 

If only one brane has positive tension, 
the discussion of the gauge hierarchy problem is similar to that in just above paragraph.

\subsection{Models on the Space-Time $M^4\times R^1/Z_2$} 
Now, we consider the models on the space-time $M^4\times R^1/Z_2$. Similar to the section 
2, 
the models with $Z_2$ symmetry and odd number of 3-branes in the subsection 3.1
 can be generalized to the models in which the fifth dimension is $R^1/Z_2$.
 One just introduces the equivalence classes: $ y \sim - y$
 and $i-th ~brane \sim  (-i)-th ~brane$, and then, modules the equivalence classes. 
 Noticing that the brane tension $V_0$ is half of its original value, i. e., 
$V_0= 3 k_0 M_X^3$, we obtain the sum of the
brane tensions is positive, too.
And we will have $m+1$ brane with positions $y_0=0 < y_1 < y_2 < ... < y_m < +\infty $.

Assuming that $k_T= \sum_{j=1}^m k_j$, we obtain 
the general solution to  $\sigma (y)$ is:
\begin{equation}
\sigma (y) = \sum_{i=0}^m k_i |y-y_i| + k_T y + c 
~.~\,
\end{equation}

And the corresponding 4-dimensional Planck scale  is:
\begin{equation}
M_{pl}^2 =  M_X^3 \left(T_{m, +\infty} + \sum_{i=0}^{m-1} T_{i, i+1} \right) 
 ~,~ \,
\end{equation}
where 
\begin{equation}
T_{m, +\infty} = {1 \over\displaystyle {2 \chi_{+\infty}}} e^{-2 \sigma (y_{m})}
 ~,~ \,
\end{equation}
if $\chi_{i, i+1} \neq 0$, then
\begin{equation}
T_{i, i+1} = {1 \over\displaystyle {2 \chi_{i, i+1}}} \left( e^{-2 \sigma (y_{i+1})}
 - e^{-2 \sigma (y_i)} \right)
 ~,~ \,
\end{equation}
and if $\chi_{i, i+1} =0$, then
\begin{equation}
T_{i, i+1} = (y_{i+1}- y_i) e^{-2  \sigma (y_i)}
 ~,~ \,
\end{equation}
where
\begin{equation}
\chi_{+\infty} = \sum_{j=0}^m k_j + k_T
 ~,~ \,
\end{equation}
\begin{equation}
\chi_{i, i+1} =
\sum_{j=i+1}^m k_j - \sum_{j=0}^i k_j - k_T
 ~.~ \,
\end{equation}
By the way, one can easily prove that $T_{i, i+1}$ is positive, which makes sure that
the 4-dimensional Planck scale is positive.

In addition, the four-dimensional GUT scale on i-th brane $M_{GUT}^{(i)}$ is
related to the five-dimensional GUT scale on i-th brane $M5_{GUT}^{(i)}$:
\begin{equation}
M_{GUT}^{(i)} = M5_{GUT}^{(i)} e^{-\sigma (y_i)} \equiv M_X e^{-\sigma (y_i)}
 ~.~ \,
\end{equation}

Now, we  discuss the simple model. 
The model with two 3-branes is equivalent to the model with three
3-branes and with $Z_2$ symmetry in the subsection 3.1, so, we donot rediscuss it here. 
We give a model with three  3-branes:
 the branes' positions are: $y_0=0$, $y_1$, $y_2$, respectively, and  the values of the 
brane 
tensions divided by $6 M_X ^3$ are: $ {1\over 2}  k_0$, $k_1$, $k_2$, respectively. 
Therefore, noticing
$k_T =   k_1 + k_2 $, we obtain:
\begin{equation}
\sigma (y) = \sum_{i=0}^{2}  k_i |y-y_i| + (k_1 + k_2 ) y +  c 
~,~\,
\end{equation}
\begin{equation}
\sigma (y_0) = k_1 y_1 + k_2  y_2 +  c 
~,~\,
\end{equation}
\begin{equation}
\sigma (y_1) = k_0 y_1 +k_2 ( y_2 - y_1) + (k_1 + k_2) y_1 + c 
~,~\,
\end{equation}
\begin{equation}
\sigma (y_2) = k_0 y_2 +k_1 ( y_2 - y_1) + (k_1 + k_2) y_2 + c 
~.~\,
\end{equation}

The four-dimensional Planck scale is:
\begin{eqnarray}
M_{pl}^2 &=& {{M_X^3} \over 2}  \left( 
- {1\over\displaystyle {  k_0 }} (  e^{-2\sigma(y_1)}-e^{-2\sigma(y_0)})
- {1\over\displaystyle { k_0 + 2 k_1 }} (  e^{-2\sigma(y_2)}-e^{-2\sigma(y_1)})
 \right.\nonumber\\&&\left.
+ {1\over\displaystyle {k_0 + 2 k_1 + 2 k_2}} e^{-2\sigma(y_2)}
 \right)
~.~\,
\end{eqnarray}

If there exist at least two branes which have positive tensions. Without loss of 
generality, we assume
the brane
with position $y_0$ has positive tension, $ \sigma(y_0) <  \sigma(y_1)$,
and $ \sigma(y_0) <  \sigma(y_2)$. If $ e^{-2(\sigma(y_1) -  \sigma(y_0))} << 1$
and $ e^{-2(\sigma(y_2) -  \sigma(y_0))} << 1$, we obtain
\begin{equation}
M_{pl}^2 = {{M_X^3}\over\displaystyle {2k_0}} e^{-2\sigma(y_0)}
~.~\,
\end{equation}
If  the brane with position $y_j$ where $j=1$ or $2$ is the observable brane,
 we can solve the gauge hierarchy problem.  
Assuming $M_X = 2 k_0 $, we obtain:
\begin{equation}
M_{GUT}^{(j)} = M_{pl} e^{-(\sigma(y_j) -  \sigma(y_0))} 
 ~.~ \,
\end{equation}
So, we can push the GUT scale in our world to TeV scale and
$10^5$ GeV scale range if  $ \sigma(y_j) -  \sigma(y_0) $ = 34.5 and 30, 
respectively. And the value of $\sigma(y_0)$ determines the
relation between $M_{pl}$ and $M_X$.

If the brane with position $y_0$  is the observable brane,
 the gauge hierarchy problem can be solved only if $\sigma(y_0) >0 $. Assuming that
$M_{pl} = 2 k_0 $ and  $e^{-2 (\sigma(y_j)-\sigma(y_0))} < < 1$ where $j=1, 2$, we obtain:
\begin{equation}
M_{GUT}^{(0)} = M_{pl} e^{-{1\over 3} \sigma(y_0)  } 
 ~,~ \,
\end{equation}
with $\sigma (y_0) $ = 103.5 and 90, we can have the GUT scale in
our world at TeV scale and $10^5$ GeV scale, respectively.
The five-dimensional Planck scale is $10^{48}$ GeV and $10^{44}$ GeV, respectively. 

For just one brane with positive brane tension case, 
 the discussion of the gauge hierarchy problem is similar to that in just above paragraph.

\subsection{Models on the Space-Time $M^4\times S^1$} 
In this subsection, we consider the models on the space-time $M^4\times S^1$.
Assuming we have $l+m+1$ parallel 3-branes, and
their fifth coordinastes are: $- \pi \rho \leq y_{-l} < y_{-l+1} < ...< y_{-1} < y_0
< y_{1} < ... < y_{m-1} < y_{m} < \pi \rho$.  The Lagrangian, the Einstein equation and
the differential equations of $\sigma(y)$ are similar to those in subsection 3.1, except 
we require that
$ |y| \leq \pi \rho$,  $\sigma(-\pi \rho) = \sigma (\pi \rho)$, and the
$\Lambda(y)$ is defined as the following:
\begin{eqnarray}
\Lambda (y) &=& \sum_{i=-l+1}^m \Lambda_i \left(\theta (y-y_{i-1}) - \theta (y-y_i) \right)
+ \Lambda_{*} (\theta (y-y_m)  
 \nonumber\\&&
+ \theta ( -y + y_{-l} ) )
~.~\,
\end{eqnarray}

The general solution to  the differential equations of $\sigma(y)$ is:
\begin{equation}
\sigma (y) = \sum_{i=-l}^m k_i |y-y_i| + k_c y + c 
~.~\,
\end{equation}

The relations between the $k_i$ and $V_i$, 
and the relations between the $k_i$  and $\Lambda_i$ are:
\begin{equation}
V_i= 6 k_i M_X^3
~,~\,
\end{equation}
\begin{equation}
\Lambda_i= -6 M_X^3 (\sum_{j=i}^m k_j - \sum_{j=-l}^{i-1} k_j-k_c)^2
~,~\,
\end{equation}
\begin{equation}
\Lambda_{*}= -6 M_X^3 k_c^2 ~,~\,
\end{equation}

${{d\sigma(y)}\over\displaystyle {dy}}$ is a piece-wise continuous function or
sectionally continuous function for it has finite jumps because of finite branes, and the 
jump
is $2k_i$ when it pass the $i-th$ brane. In fact, ${{d\sigma(y)}\over\displaystyle {dy}}$ 
is
sectional constant. In addition, 
 if starting a point not belong to any brane,  we wind around the circle one time,
the sum of the total jumps for the function ${{d\sigma(y)}\over\displaystyle {dy}}$
is $2 \sum_{i=-l}^{m} k_i$ which is equal to zero because ${{d\sigma(y)}\over\displaystyle 
{dy}}$ 
is sectional constant. Therefore, we obtain the sum of the brane tensions should be zero: 
\begin{equation}
\sum_{j=-l}^{m} k_j  = 0~.~\,
\end{equation} 
 
In addition, from $\sigma(-\pi \rho) = \sigma (\pi \rho)$, we obtain the constraint:
\begin{equation}
\sum_{j=-l}^{m} k_j y_j = k_c \pi \rho~.~\,
\end{equation}

And the corresponding 4-dimensional Planck scale  is:
\begin{equation}
M_{pl}^2 =  M_X^3 \left(  T_{m, -l} + \sum_{i=-l}^{m-1} T_{i, i+1} \right) 
 ~,~ \,
\end{equation}
where if $k_c \neq 0$, then
\begin{equation}
T_{m, -l} = -{1 \over\displaystyle {2 k_c}} \left( e^{-2 \sigma (y_{-l})} -
 e^{-2 \sigma (y_{m})}\right)
 ~,~ \,
\end{equation}
and if $k_c = 0$, then
\begin{equation}
T_{m, -l} =  \left( 2 \pi \rho + y_{-l} - y_m \right) e^{-2 \sigma (y_{-l})}
 ~,~ \,
\end{equation}

if $\chi_{i, i+1} \neq 0$, then
\begin{equation}
T_{i, i+1} = {1 \over\displaystyle {2 \chi_{i, i+1}}} \left( e^{-2 \sigma (y_{i+1})}
 - e^{-2 \sigma (y_i)} \right)
 ~,~ \,
\end{equation}
and if $\chi_{i, i+1} =0$, then
\begin{equation}
T_{i, i+1} = (y_{i+1}- y_i) e^{-2  \sigma (y_i)}
 ~,~ \,
\end{equation}
where
\begin{equation}
\chi_{i, i+1} =
\sum_{j=i+1}^m k_j - \sum_{j=-l}^i k_j - k_c
 ~.~ \,
\end{equation}
As before, one can easily prove that $T_{i, i+1}$ is positive, which makes sure that
the 4-dimensional Planck scale is positive.

In addition, the four-dimensional GUT scale on i-th brane $M_{GUT}^{(i)}$ is
related to the five-dimensional GUT scale on i-th brane $M5_{GUT}^{(i)}$:
\begin{equation}
M_{GUT}^{(i)} = M5_{GUT}^{(i)} e^{-\sigma (y_i)} \equiv M_X e^{-\sigma (y_i)}
 ~.~ \,
\end{equation}

The models can be generalized to the models 
with $Z_2$ symmetry. There are four kinds of such models, two  kinds of such models have 
odd number
of the branes, the other two have even number of the branes. 
For the models with odd number of the branes: (I) one requires that: 
 $k_{-i} = k_i$, $ y_{-i} = - y_i$,
 $m=l$ and $k_c=0$.   (II)  one requires that: $k_{-i} = k_i$, $y_{-i} = -y_i$ for $1 \leq 
i \leq m$,
 $l=m+1$, $k_0 = 0$ (no number 0 brane), $y_{-l}=-\pi \rho$ and $k_c=-k_{-l}$. For the 
models with even
 number of the branes; (I) one  requires that $k_{-i} = k_i$, $y_{-i} = -y_i$,
 $m=l$, $k_0=0$ (no number 0 brane) and $k_c=0$. (II) one  requires that
$k_{-i} = k_i$, $y_{-i} = -y_i$, for $1 \leq i \leq m$,
 $l=m+1$, $k_0 \ne 0$, $y_{-l}=-\pi \rho$ and $k_c=-k_{-l}$. 
 
Let us discuss two  simple models. 

(I) Model with three 3-branes: 
the branes' positions are $y_1$, $y_2$, $y_3$, respectively, and  the values of the brane 
tensions divided by $6 M_X ^3$ are: $ k_1$, $-(k_1+k_3)$, $k_3$, respectively, where 
we assume $k_1 >0$, $k_3 > 0$.
Therefore, we obtain:
\begin{equation}
\sigma (y) = k_1 |y-y_1|-(k_1+k_3) |y-y_2| + k_3 | y-y_3 | + k_c y + c 
~.~\,
\end{equation}
The constraint equation is
\begin{equation}
 k_1 y_1-(k_1+k_3) y_2 + k_3 y_3  =  k_c \pi \rho
~.~\,
\end{equation}
and
\begin{equation}
\sigma (y_1) = k_c (  \pi \rho + y_1 ) + c 
~,~\,
\end{equation}
\begin{equation}
\sigma (y_2) = 2 k_1 (y_2 -y_1) + k_c ( \pi \rho + y_2) + c 
~,~\,
\end{equation}
\begin{equation}
\sigma (y_3) = k_c (y_3 - \pi \rho) + c 
~.~\,
\end{equation}
Without loss of generality, assuming $ k_c > 0$, we obtain 
$\sigma(y_3) < \sigma (y_1) < \sigma (y_2) $.

The four-dimensional Planck scale is
\begin{eqnarray}
M_{pl}^2 &=& {{M_X^3} \over 2}  \left( - {1\over\displaystyle { k_c}} ( e^{-2\sigma(y_1)} - 
e^{-2\sigma(y_3)})
- {1\over\displaystyle { 2 k_1 + k_c }} (  e^{-2\sigma(y_2)}-e^{-2\sigma(y_1)})
 \right.\nonumber\\&&\left.
+ {1\over\displaystyle {2 k_3- k_c  }} (  e^{-2\sigma(y_3)}-e^{-2\sigma(y_2)})
 \right)
~.~\,
\end{eqnarray}

 If $ e^{-2(\sigma(y_1) -  \sigma(y_3))} << 1$,  we obtain:
\begin{equation}
M_{pl}^2 = {M_X^3} {{ k_3}\over\displaystyle {(2k_3 -k_c) k_c}} e^{-2 k_c (y_3 - \pi \rho) 
- 2 c}
~.~\,
\end{equation}
Because of the constraint equation, $ 2k_3 -k_c > 0$.
If  the brane with position $y_1$  is the observable brane, we 
can solve the gauge hierarchy problem.  Assuming
$M_X = {{(2k_3 -k_c) k_c} \over\displaystyle {k_3}} $, we obtain:
\begin{equation}
M_{GUT}^{(1)} = M_{pl} e^{- k_c (2 \pi \rho + y_1 -y_3)} 
 ~.~ \,
\end{equation}
So, we can push the GUT scale in our world to TeV scale and
$10^5$ GeV scale range if  $ k_c (2 \pi \rho + y_1 -y_3) $ = 34.5 and 30, 
respectively. And the value of $\sigma(y_3)$ determines the
relation between $M_{pl}$ and $M_X$.

If the brane with position $y_3$  is the observable brane,
 we can solve the gauge hierarchy problem only if $\sigma(y_3) >0 $. Assuming that
$M_{pl} = {{(2k_3 -k_c) k_c} \over\displaystyle {k_3}} $ and
$ e^{-2(\sigma(y_1) -  \sigma(y_3))} $ $<$ $ <$ $ 1$, we obtain:
\begin{equation}
M_{GUT}^{(3)} = M_{pl} e^{-{1\over 3} \sigma(y_3)  } 
 ~,~ \,
\end{equation}
with $\sigma (y_3) $ = 103.5 and 90, we can have the GUT scale in
our world at TeV scale and $10^5$ GeV scale, respectively.
The five-dimensional Planck scale is 
$10^{48}$ GeV and $10^{44}$ GeV, respectively. 

(II) Model with four 3-branes and $Z_2$ symmetry. 
The branes' positions are $y_{-2}= -\pi \rho $, $y_{-1} = -y_1$, $y_0 = 0$, $y_1$, 
respectively, and  the values of 
the brane 
tensions divided by $6 M_X ^3$ are: $ 2 k_2$, $ -(k_0+k_2)$, $2 k_0$, $-(k_0 + k_2)$, 
respectively, where
we assume $k_0 > 0, k_2 > 0$. Therefore, we obtain:
\begin{equation}
\sigma (y) = -(k_2+ k_0) (|y-y_1|+ |y+y_1|)+ 2 k_0 |y| + c 
~,~\,
\end{equation}
then,
\begin{equation}
\sigma (y_{-2}) = - 2 k_2 \pi \rho + c ~,~ \sigma (y_{0}) = - 2 (k_2 + k_0) y_1 + c 
~,~\,
\end{equation}
\begin{equation}
\sigma (y_{-1}) = \sigma (y_1) = -2 k_2 y_1 + c
~.~\,
\end{equation}

The four-dimensional Planck scale is
\begin{eqnarray}
M_{pl}^2 &=& {{M_X^3}\over 4}  \left(-{1\over k_2} ( e^{-2 \sigma (y_{-1})} - e^{-2 \sigma 
(y_{-2})} )
+ {1\over k_0} ( e^{-2 \sigma (y_{0})} - e^{-2 \sigma (y_{-1})} )
 \right.\nonumber\\&&\left.
- {1\over k_0} ( e^{-2 \sigma (y_{1})} - e^{-2 \sigma (y_{0})} )
+ {1\over k_2} ( e^{-2 \sigma (y_{-2})} - e^{-2 \sigma (y_{1})})
\right)
~.~\,
\end{eqnarray}
Without loss of generality, we assume that $ \sigma (y_{-2}) < \sigma (y_{0})$ or 
$k_2 \pi \rho - (k_2 + k_0) y_1 > 0$.
If  the brane with position $y_{0}$  is the observable brane, 
 the gauge hierarchy problem can be solved. Assuming that  
$ e^{-2(\sigma(y_{0}) -\sigma (y_{-2}))} < < 1$,
and $M_X=2 k_2$, one obtains:
\begin{equation}
M_{GUT}^{(0)} = M_{pl} e^{-(\sigma(y_{0}) -\sigma (y_{-2}))} 
 ~.~ \,
\end{equation}
So, one can push the GUT scale in our world to TeV scale and
$10^5$ GeV scale range if  $\sigma(y_{0}) -\sigma (y_{-2}) = 2( k_2 \pi \rho - (k_2 + k_0) 
y_1) =$ 34.5 and 30, 
respectively. And the value of $\sigma (y_{-2})$ determines the
relation between $M_{pl}$ and $M_X$.

\subsection{Models on the Space-Time $M^4\times S^1/Z_2$}
The models with $Z_2$ symmetry in the last subsection 3.3 
 can be generalized to the models in which the fifth dimension is $S^1/Z_2$.
 One just requires that $k_{-i} = k_i$, $ y_{-i} = - y_i$, for $1 \leq i \leq m$,
 $l=m+1$,  $y_{-l} = -\pi \rho$, and $k_c = - k_{-l}$, then, introduces the equivalence 
classes: $ y \sim - y$
 and $i-th ~brane \sim  (-i)-th ~brane$ for $i=1, ..., m$.
 After moduling the eqivalence classes, we obtain the models on  $M^4\times S^1/Z_2$. We 
renumber
 $(-l)-th$ brane as $(m+1)-th$ brane, so, $y_{m+1} = \pi \rho$. Using splitting brane 
method, we obtain 
 the brane tensions $V_0$ and $V_{m+1}$ are half of their original values, i. e., 
$V_0= 3 k_0 M_X^3$, $V_{m+1}= 3 k_{m+1} M_X^3$. Of course,
the sum of the brane tensions is zero, too, i.e., $ k_0 + k_{m+1} + 2\sum_{i=1}^m k_i =0$.
And we will have $m+2$ branes with positions
$y_0=0 < y_1 <... < y_m < y_{m+1}=\pi \rho$.  

The general solution of $\sigma(y)$ is:
\begin{equation}
\sigma (y) =  k_0 |y| + \sum_{i=1}^m k_i (|y-y_i| + |y+ y_i|) + c 
~.~\,
\end{equation}
And we can rewrite it as
\begin{equation}
\sigma (y) =   \sum_{i=1}^m k_i |y-y_i| + k_T y + c 
~,~\,
\end{equation}
where $k_T = {1\over 2} (k_0 - k_{m+1})$.

The corresponding 4-dimensional Planck scale  is:
\begin{equation}
M_{pl}^2 =  M_X^3 \left(   \sum_{i=0}^{m} T_{i, i+1} \right) 
 ~,~ \,
\end{equation}
where 
if $\chi_{i, i+1} \neq 0$, then
\begin{equation}
T_{i, i+1} = {1 \over\displaystyle {2 \chi_{i, i+1}}} \left( e^{-2 \sigma (y_{i+1})}
 - e^{-2 \sigma (y_i)} \right)
 ~,~ \,
\end{equation}
and if $\chi_{i, i+1} =0$, then
\begin{equation}
T_{i, i+1} = (y_{i+1}- y_i) e^{-2  \sigma (y_i)}
 ~,~ \,
\end{equation}
where
\begin{equation}
\chi_{0, 1} = - k_0 ~,~ \chi_{m, m+1} = k_{m+1}
~,~ \,
\end{equation}
and for $i=1, ..., m-1,$
\begin{equation}
\chi_{i, i+1} =
\sum_{j=i+1}^m k_j - \sum_{j=1}^i k_j - k_T
 ~.~ \,
\end{equation}
By the way, one can easily prove that $T_{i, i+1}$ is positive, which makes sure that
the 4-dimensional Planck scale is positive.

In addition, the four-dimensional GUT scale on i-th brane $M_{GUT}^{(i)}$ is
related to the five-dimensional GUT scale on i-th brane $M5_{GUT}^{(i)}$:
\begin{equation}
M_{GUT}^{(i)} = M5_{GUT}^{(i)} e^{-\sigma (y_i)} \equiv M_X e^{-\sigma (y_i)}
 ~.~ \,
\end{equation}

Now, we discuss the simple model. 
The model with three 3-branes is equivalent to the model with four 3-branes and with $Z_2$
symmetry in the last subsection 3.3. So, we will not rediscuss it here. We discuss the 
model with four  
3-branes: the branes' positions are $y_0=0$, $y_1$, $y_2$, $y_3=\pi \rho$, respectively, 
and  
the values of the brane 
tensions divided by $6 M_X ^3$ are: ${1\over 2}(-k_1 + k_2 +k_T)$,
 $ k_1 $, $-k_2 $, $-{1\over 2}( k_1-k_2 + k_T)$, respectively, where 
 we assume $k_1 > 0, k_2 > 0$. Therefore, we obtain:
\begin{equation}
\sigma (y) = k_T y +   k_1 |y-y_1| - k_2 |y-y_2| + c 
~,~\,
\end{equation}
\begin{equation}
\sigma (y_0) = k_1 y_1 - k_2 y_2 + c ~,~\,
\end{equation}
\begin{equation}
\sigma (y_1) =- k_2 ( y_2-y_1) + k_T y_1 +c
~,~\,
\end{equation}
\begin{equation}
\sigma (y_2) = k_1 (y_2 - y_1) +k_T y_2 +  c ~,~\,
\end{equation}
\begin{equation}
 \sigma (y_3) =(k_1-k_2+k_T) \pi \rho- (k_1 y_1-k_2 y_2) + c
~.~\,
\end{equation}
\begin{eqnarray}
M_{pl}^2 &=& -{{M_X^3}\over 2}  \left({1\over\displaystyle {k_2 + k_T -k_1}} 
( e^{-2 \sigma (y_{1})} - e^{-2 \sigma (y_{0})} )
 \right.\nonumber\\&&\left.
+ {1\over\displaystyle { k_1 + k_2 + k_T} } ( e^{-2 \sigma (y_{2})} - e^{-2 \sigma (y_{1})} 
)
 \right.\nonumber\\&&\left.
+ {1\over\displaystyle { k_1 - k_2 + k_T}  } ( e^{-2 \sigma (y_{3})} - e^{-2 \sigma 
(y_{2})} )
\right)
~.~\,
\end{eqnarray}
Without loss of generality, we assume that $k_T > 0$ and $k_2 + k_T > k_1$. Then, the brane 
at $y_0$ has 
positive tension and
 $\sigma (y_0) < \sigma (y_1)$. Assuming $ e^{-2 (\sigma (y_{j})-\sigma (y_{0})) } << 1$ 
 for $j=1, 2, 3$,  $M_X= 2 ( k_2 + k_T-k_1)$, 
and the brane with position $y_1$  is the observable brane,
 we can solve the gauge hierarchy problem: 
\begin{equation}
M_{GUT}^{(1)} = M_{pl} e^{-(k_2-k_1+k_T) y_1} 
 ~.~ \,
\end{equation}
So, we can push the GUT scale in our world to TeV scale and
$10^5$ GeV scale range if  $ (k_2-k_1+k_T) y_1 $ = 34.5 and 30, 
respectively. And the value of $\sigma(y_0)$ determines the
relation between $M_{pl}$ and $M_X$.

If the brane with position $y_0$  is the observable brane,
 one can solve the gauge hierarchy problem only if $\sigma(y_0) >0 $, 
i. e., $k_1 y_1 > k_2 y_2$. Assuming that
$M_{pl} = 2 ( k_2 + k_T-k_1)  $ and
$ e^{-2 (k_2-k_1+k_T) y_1 } << 1$, one obtains:
\begin{equation}
M_{GUT}^{(0)} = M_{pl} e^{-{1\over 3} \sigma(y_0)  } 
 ~,~ \,
\end{equation}
with $\sigma (y_0) $ = 103.5 and 90, one can push the GUT scale in
our world to TeV scale and $10^5$ GeV scale range, respectively.
The five-dimensional Planck scale is $10^{48}$ GeV and $10^{44}$ GeV, respectively.

\section{Conclusion}

If the fifth dimension is one-dimensional connected manifold or one-dimensional connected
manifold with boundary, up to diffeomorphic, the only possible
space-time will be $M^4 \times R^1$, $M^4 \times R^1/Z_2$,
$M^4 \times S^1$ and $M^4 \times S^1/Z_2$. 
And there exist two 
possibilities on cosmology constant along the fifth dimension: the cosmology
constant is constant, and the cosmology constant
is sectional constant.
For any point in 
$M^4 \times R^1$, $M^4 \times R^1/Z_2$, $M^4 \times S^1$ and $M^4 \times S^1/Z_2$,
 which is not belong to any brane and the section where the cosmology constant is zero, 
there is a neighborhood which is diffeomorphic to ( or a slice of ) $AdS_5$ space.
We construct the general models with  parallel 3-branes 
and with constant/sectional constant cosmology
constant along the fifth dimenison
on the space-time $M^4 \times R^1$, $M^4 \times R^1/Z_2$,
$M^4 \times S^1$ and $M^4 \times S^1/Z_2$, and point out that for compact fifth dimension, 
the sum of the brane tensions
is zero, for non-compact fifth dimension, the sum of the brane tensions is positive if
we require that the 4-dimentional Planck scale is finite.
We assume the observable brane which includes our world should have
positive tension, and conclude that
in those general scenarios, the 5-dimensional GUT scale on each brane
can be indentified as the 5-dimensional Planck scale, but,
the 4-dimensional Planck scale is generated from the low 4-dimensional
GUT scale exponentially in our world. 
We also give some simple models
to show explicitly how to solve the gauge hierarchy problem.

\section*{Note Added}
During the preparation of this paper, ref.~\cite{NEWNK} appeared, which overlapped the part
of this paper. In ref.~\cite{NEWNK}, the author derived the crystal braneworld solutions, 
comprising of intersecting families of parallel $n+2$-branes in a $4+n$-dimensional $AdS$
space with the constant cosmology constant in the whole space-time. Therefore, the models
in section 2 may be considered as specific models in ref.~\cite{NEWNK}, although we discuss
the models in detail. However, we also discuss the models with sectional constant cosmology
constant along the fifth dimension in section 3. By the way, when we construct the models
with many brane intersections/junctions on the space-time $M^4 \times M^n$, the cosmology 
constant
might not be constant in the whole space-time, and $M^n$ should be n-dimensional manifold 
or n-dimensional manifold with boundary~\cite{LTJII}.

\section*{Acknowledgments}
I would like to thank S. Nam for bringing my attention to his paper.
This research was supported in part by the U.S.~Department of Energy under
 Grant No.~DE-FG02-95ER40896 and in part by the University of Wisconsin 
 Research Committee with funds granted by the Wisconsin Alumni
  Research Foundation.


\newpage

\begin{thebibliography}{99}
\itemsep 0.5mm
\bibitem{AADD} N. Arkani-Hamed, S. Dimopoulos and G. Dvali, Phys. Lett. 
B{\bf 429} (1998) 263, hep-ph/9803315; I. Antoniadis, 
N. Arkani-Hamed, S. Dimopoulos and G. Dvali, Phys. Lett. B{\bf 436}
 (1998) 257,
 hep-ph/9804398.
\bibitem{LRRS} L. Randall and R. Sundrum, hep-ph/9905221.
\bibitem{LRRSN} L. Randall and R. Sundrum, hep-ph/9906064.
\bibitem{JLLR} J. Lykken and L. Randall, hep-th/9908076.
\bibitem{MG} M. Gogberashvili: hep-ph/9812296, hep-ph/9812365,
hep-ph/9904383, hep-ph/9908347.
\bibitem{HDDK} N. Arkani-Hamed, S. Dimopoulos, G. Dvali and N. Kaloper,
hep-th/9907209.
\bibitem{MBW} W. D. Goldberger and M. B. Wise, hep-ph/9907447, hep-ph/9911457.
\bibitem{TN} T. Nihei, hep-ph/9905487.
\bibitem{CGKT} C. Csaki, M. Grasesser, C. Kolda and T. Terning,
hep-ph/9906513.
\bibitem{NK} N. Kaloper, hep-th/9905210.
\bibitem{HV} H. Verlinde, hep-th/9906182.
\bibitem{IO} I. Oda, hep-th/9908104.
\bibitem{ABKS} A. Brandhuber and K. Sfetsos, hep-th/9908116.
\bibitem{AK} A. Kehagias, hep-th/9908174. 
\bibitem{HBDK} H. B. Kim and H. D. Kim, hep-th/9909053.
\bibitem{TSMS} T. Shiromizu, K. Maeda, M. Sasaki, gr-qc/9910076.
\bibitem{CVETIC} K. Behrndt and M. Cvetic, hep-th/9909058.
\bibitem{LTJ} T. Li, hep-th/9908174, Phys. Lett. B to appear.
\bibitem{LTJI} T. Li, hep-th/9911234.
\bibitem{LTJII} T. Li, in preparation.
\bibitem{KRDDG} K. R. Dienes, E. Dudas, and T. Gherghetta, hep-ph/9908530.
\bibitem{DWFGK} O. DeWolfe, D.Z. Freedman, S.S. Gubser and A. Karch, hep-th/9909134.
\bibitem{CY}  S. Chang and M. Yamaguchi, hep-ph/9909523.
\bibitem{CP}  A. Chodos and E. Poppit,  hep-th/9909199.
\bibitem{GCS}  C. Grojean, J. Cline and G. Servant, hep-th/9906523, hep-ph/9909496,
hep-th/9910081.
\bibitem{CGRT} C. Csaki, M. Grasesser, L. Randall and T. Terning,
hep-ph/9911406, and refernce therein.
\bibitem{CCYS}  C. Csaki and Y. Shirman, hep-th/9908186.
\bibitem{AEN}A. E. Nelson, hep-th/9909001.
\bibitem{ODA} I. Oda, hep-th/9909048.
\bibitem{HSTT} H. Hatanaka, M. Sakamoto, M. Tachibana, and K. Takenaga,
hep-th/9909076.
\bibitem{NEWNK} N. Kaloper, hep-th/9912125.
\bibitem{SNAM} S. Nam, hep-th/9911237.
\bibitem{MCHS} M. Cvetic and H. H. Soleng, Phys. Rept. {\bf 282}, 159 (1997)
hep-th/9604090, and references therein.
\bibitem{GUTP}  K.R. Dienes, E. Dudas, and T. Gherghetta, hep-ph/9803466, 
hep-ph/9806292, hep-ph/9807522. 


\end{thebibliography}
\end{document}